\documentclass[amsmath,amssymb,aps,citeautoscript,superscriptaddress,twocolumn,showpacs,10pt]{revtex4-1}

\usepackage{bm}

\usepackage{graphicx}
\usepackage{dcolumn}
\usepackage[colorlinks,linkcolor=red,anchorcolor=green,
citecolor=blue,breaklinks]{hyperref}

\usepackage{mathrsfs}
\usepackage{color}
\usepackage{braket}
\usepackage[normalem]{ulem}
\usepackage{algorithm}
\usepackage{algorithmicx}
\usepackage{algpseudocode}
\usepackage[normalem]{ulem}

\newcommand{\add}[1]{{\color{red}#1}}

\graphicspath{{figures/}}
\newcommand{\mK}{\mathcal{K}}

\newcommand{\hA}{\hat{A}}

\newcommand{\hP}{\hat{P}}

\usepackage{qcircuit} 
 
\begin{document}

\preprint{}

\title{Quantum Davidson Algorithm for Excited States 
}

\affiliation{Theoretical Division, Los Alamos National Laboratory, Los Alamos, NM 87545, USA}
\affiliation{Chemistry Division, Los Alamos National Laboratory, Los Alamos, NM 87545, USA}
\affiliation{Department of Chemistry and Biochemistry, Utah State University, Logan, Utah 84322, USA}

\author{Nikolay V. Tkachenko}
\altaffiliation[Now at ]{College of Chemistry, University of California, Berkeley}
\affiliation{Theoretical Division, Los Alamos National Laboratory, Los Alamos, NM 87545, USA}
\affiliation{Department of Chemistry and Biochemistry, Utah State University, Logan, Utah 84322, USA}

\author{Lukasz Cincio}
\affiliation{Theoretical Division, Los Alamos National Laboratory, Los Alamos, NM 87545, USA}

\author{Alexander I. Boldyrev}
\affiliation{Department of Chemistry and Biochemistry, Utah State University, Logan, Utah 84322, USA}

\author{Sergei Tretiak}
\affiliation{Theoretical Division, Los Alamos National Laboratory, Los Alamos, NM 87545, USA}

\author{Pavel A. Dub}
\email{pavel.dub@schrodinger.com}
\altaffiliation[Now at ]{Schr\"odinger, Inc.}
\affiliation{Chemistry Division, 
Los Alamos National Laboratory, Los Alamos, NM 87545, USA}

\author{Yu Zhang}
\email{zhy@lanl.gov}
\affiliation{Theoretical Division, Los Alamos National Laboratory, Los Alamos, NM 87545, USA}

\date{\today}

\begin{abstract}
Excited state properties play a pivotal role in various chemical and physical phenomena, such as charge separation and light emission. However, the primary focus of most existing quantum algorithms has been the ground state, as seen in quantum phase estimation and the variational quantum eigensolver (VQE). Although VQE-type methods have been extended to explore excited states, these methods grapple with optimization challenges. In contrast, the quantum Krylov subspace (QKS) method has been introduced to address both ground and excited states, positioning itself as a cost-effective alternative to quantum phase estimation. However, conventional QKS methodologies depend on a pre-generated subspace through real or imaginary-time evolutions. This subspace is inherently expansive and can be plagued with issues like slow convergence or numerical instabilities, often leading to relatively deep circuits. Our research presents an economic QKS algorithm, which we term the quantum Davidson (QDavidson) algorithm. This innovation hinges on the iterative expansion of the Krylov subspace and the incorporation of a pre-conditioner within the Davidson framework. By using the residues of eigenstates to expand the Krylov subspace, we manage to formulate a compact subspace that aligns closely with the exact solutions. This iterative subspace expansion paves the way for a more rapid convergence in comparison to other QKS techniques, such as the quantum Lanczos. Using quantum simulators, we employ the novel QDavidson algorithm to delve into the excited state properties of various systems, spanning from the Heisenberg spin model to real molecules. Compared to the existing QKS methods, the QDavidson algorithm not only converges swiftly but also demands a significantly shallower circuit. This efficiency establishes the QDavidson method as a pragmatic tool for elucidating both ground and excited state properties on quantum computing platforms.
\end{abstract}

\maketitle

\section{Introduction} 
Computing the ground and excited state properties of intricate many-body systems is a cornerstone of quantum physics and chemistry. Despite the importance, this endeavor demands substantial computational power due to the factorial growth of the full many-body wavefunction's solution space as the system size (represented by the number of electrons and basis functions)~\cite{cao2018quantum,mcardle2020quantum,garnet2020cr}. As a result, classical quantum chemistry techniques such as Hartree-Fock (HF), density functional theory (DFT)~\cite{weitao2012cr,dft2009natphys}, tensor network methods~\cite{dmrgrev2011, SCHOLLWOCK201196} that optimize the wavefunction in the form of a matrix product states (MPS), selected configuration interaction (sCI)~\cite{jctc.8b00536, jctc.9b00476} that iteratively expands the configuration interaction (CI) spaces, and coupled-cluster (CC) theory truncated at finite orders~\cite{cr2001417} have been conceived to bypass the direct formulation of full many-body wavefunctions. However, these techniques invariably employ truncations or approximations and are limited to a certain size.

Since the 1980s, quantum computers (QC) that leverage the power of quantum entanglement have been proposed as the ideal platforms for simulating quantum mechanics~\cite{RN131, manin}. Advancing into the Noisy Intermediate-Scale Quantum (NISQ) era~\cite{preskil2018}, quantum computing has shown its potential with the demonstration of quantum advantages in well-defined tasks~\cite{supermacy2019}. Electronic structure problems, crucial to various scientific disciplines, emerge as one of the most promising and immediate applications of quantum computers~\cite{cao2018quantum,mcardle2020quantum,garnet2020cr, PRXQuantum.2.017001}. Numerous quantum algorithms have been proposed for calculating the ground state of quantum many-body systems on QC since the advent of quantum phase estimation (QPE)~\cite{RPL83.5162,kitaev2002classical}. However, the inherently noisy nature of NISQ devices necessitates hybrid quantum-classical algorithms with shallower circuits. To this end, the Variational Quantum Eigensolver (VQE) that leverages the power of variational principle and classical optimization was conceived accordingly~\cite{RN155}. The VQE scheme deploys the parameterized wavefunction(or ansatz) and the corresponding energy measurement on QCs and then uses classical algorithms for variational energy minimization. 
VQE algorithm has been performed on various quantum architectures such as photons~\cite{RN155}, superconducting qubits~\cite{RN748, Kandala2017Nature}, and trapped ions~\cite{trapion2020}. Since then, many algorithms have been proposed to improve the performance further or reduce the quantum resource requirements of VQE~\cite{permVQE2021,jctc9b01084, Grimsley2019NatCommun,clusterVQE2021, PRXQuantum.2.020310, Ryabinkin_2021,artur2021bch,alan2021mivqe, RN170, PRXQuantum.2.020310, jctc9b01084, RN669, Ryabinkin_2021,yordanov2020iterative,hybridtensor,2021arXiv210810434G,cerezo2020natrev,bharti2021noisy}.

Despite the extensive development of quantum algorithms for electronic structure problems, the majority pivots on ground state properties. However, many chemical critical processes, including energy transfer~\cite{Nelson2019natcomm}, bond dissociation~\cite{zhang2020jctc}, and light emission~\cite{cr9b00447}, revolve around electronically excited states. Which necessitates the development of quantum algorithms for excited states. One straightforward way is to extend VQE for excited states by introducing certain constraints~\cite{PRResearch.1.033062,deflation2019,mcvqe2019,PhysRevA.95.020501,PhysRevA.95.042308,PhysRevResearch.1.033062,Higgott_2019,Kawai_2020,qua.26352}. Despite the success and impact of VQE algorithms, they suffer from optimization problems. The optimization process in VQE is challenging due to the high nonlinearity of the energy and stochastic errors~\cite{vqoa2021np} and is compounded by the inclusion of multiple excited states.

Alternatively, the other emerging direction for calculating excited states on QC is based on the quantum subspace, showcased by Quantum subspace expansion (QSE)~\cite{PhysRevA.95.042308,PhysRevX.8.011021, PRX10.011004,filter2019,qeom2020}, non-orthogonal VQE~\cite{Huggins_2020}, equation of motions~\cite{Asthana:2023uv, kumar_quantum_2023}, and the quantum Krylov subspace (QKS) framework inspired by its classical analogs~\cite{Cortes:2022uu, jctc.9b01125,GarnetNP2020,gray2022qks,travis2021benchmark,lin2021qks}. Current QKS methods utilize either real or imaginary time~\cite {jctc.9b01125,GarnetNP2020} evolutions to generate the Krylov subspace, which then is used to sample the low-lying spectrum of the Hamiltonian $\hat{H}$. In particular, the Quantum Lanczos (QLanczos) algorithm~\cite{GarnetNP2020} that engages a basis of correlated states generated from the imaginary-time propagation~\cite{RN1234, Siopsis2020} has been proposed. Even though the QKS methods remove the optimization problems of the VQE algorithms, the current QKS usually requires a relatively deep circuit due to the trotter expansion of the real/imaginary time evolution and a larger number of iterations for convergence. Moreover, the pre-generated Krylov subspace from the time evolution is not necessarily compact.

This research introduces the Quantum Davidson (QDavidson) algorithm, an economic QKS method that efficiently crafts a compact Krylov subspace. By harnessing the Residue and a pre-conditioner, Davidson's algorithm restricts the subspace search near the exact state, guaranteeing brisk convergence~\cite{saad03book}. Compared to its classical counterpart, our strategy offloads subspace expansion to the quantum computer, eliminating the explicit construction of full many-body wavefunctions on classical computers, leading to the speedup in the subspace projection. Compared to the existing QKS method, QDavidson's rapid convergence results in a much shallower circuit, making it potentially more noise resilient.

\section{Theory}

\subsection{Krylov subspace and classical Davidson algorithm}
In linear algebra, an order-$r$ Krylov subspace generated by a matrix $H$ and a reference vector $b$ is the linear subspace spanned by the images of $b$ under the first $r$ powers of $H$~\cite{lanczos1950}. This subspace is denoted as $\mathcal{K}_r(H,b) \equiv \{b, Hb, H^2b, \cdots, H^{r-1}b\}$. The Krylov subspace has been extensively utilized in numerical algorithms for finding solutions to high-dimensional matrices, such as the generalized minimal residual method (GMRES), Davidson, and quasi-minimal residual (QMR) algorithms~\cite{saad03book}.

In quantum chemistry, the iterative Krylov subspace $\mathcal{K}_r(\hat{H},\ket{\psi})$ is especially valuable for identifying low-lying states in electronic structure theory since the number of such states is typically much smaller than the size of the solution space. Various versions of the Davidson algorithm have been developed to enhance convergence by designing efficient pre-conditioners~\cite{jctc6b00459,davidson2009sergei,davidson2021shane}. Furthermore, the Davidson algorithm has been generalized to use non-orthogonal Krylov spaces~\cite{furche2016krylov}. The flowchart of a standard classical Davidson algorithm is presented in Algorithm~\ref{alg_dav0} of the Supplementary Materials (SM). Despite its computational efficiency, the generation of the subspace $\hat{H}^r\ket{\psi}$ remains a significant bottleneck in the Davidson algorithm and becomes computationally intensive on classical computers as the dimension of $\hat{H}$ grows.

\subsection{Quantum Davidson algorithm}
In this study, we introduce a quantum counterpart of the Davidson algorithm, termed QDavidson, that harnesses quantum computers to mitigate the scaling challenges of generating the Krylov subspace. Compared to the traditional Krylov method, the QKS scheme encodes arbitrarily complex states on quantum circuits, where the matrix's projection into the subspace is measured. The benefits of QKS methods over VQE-based techniques for computing low-lying excited states include 1) an independent ansatz for each reference state with a streamlined quantum circuit and 2) the elimination of the intricate optimization process. In Reference~\onlinecite{GarnetNP2020}, quantum imaginary time evolution (QITE) is conducted using the Trotter decomposition of the evolution operator $e^{-\hat{H} t}$ and mapping each Trotter step into a unitary. Subsequently, the QLanczos algorithm is introduced in the Krylov space derived from QITE snapshots. However, the QKS arising from sequential time evolution lacks compactness. Here, we develop the QDavidson algorithm to create a compact QKS set, yielding a more efficient and numerically stable realization of the QKS approach.

Instead of pre-generating the subspace through real or imaginary-time evolution, the QDavidson framework adaptively expands the QKS, keeping the subspace growth closely with the true eigenspace. Initially, orthogonal states serve as the reference. HF $\ket{\phi_0}$, single-excitation configurations $\ket{\phi^a_i}=a^\dagger_a a_i\ket{\phi_0}$, and configuration-interaction single (CIS) states $\{\sum_{ia}C_{ia}\ket{\phi^a_i}\}$, which are efficiently computed on classical computers, are natural choices for initial reference states. The set of these initial reference states is denoted as $\{\ket{\psi_k}\}$. Within the QKS framework, each basis (subsequently called a Krylov vector) within the Krylov subspace of the $j^{\text{th}}$ Davidson iteration assumes the following ansatz,
\begin{equation}\label{krylov_ref}
\ket{\psi_K} = 
 \hat{U} ^j_K \sum_{I} C_{KI} \ket{\phi_{I}}
\equiv 
 \hat{U} ^j_K \ket{\Phi_{K}}.
\end{equation}
Here, multiple reference states (i.e., a linear combination of HF/CIS states) are utilized. The entanglers $\hat{U}^j_K$ arise from the Krylov space expansion and introduce correlations that extend beyond the initial states. The details of $\hat{U}^j_K$ will be detailed later, with $\hat{U}^1_K = 1$ for the initial iteration.

\begin{figure}
  \centering
  \includegraphics[width=0.5\textwidth]{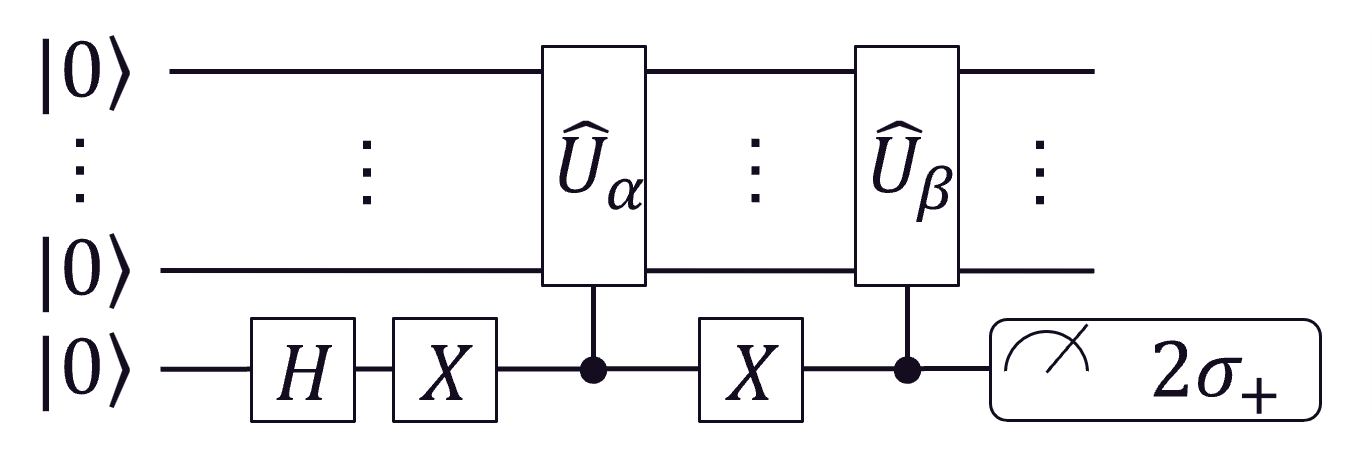}
  \caption{Schematic diagram of modified Hadamard test for measuring off-diagonal elements. The measurement of the expectation value of $2\sigma_{+} = X + iY$ operator.}
  \label{fig:htest}
\end{figure}

The general ground and excited states, denoted $\ket{\Phi_I}$, can be expressed as linear combinations of the Krylov vectors,
\begin{equation}\label{eq_refs}
\ket{\Psi_I} = \sum_{K} V_{KI} \ket{\psi_K}.
\end{equation}
Therefore, the challenge of finding the ground state and low-lying excited states, represented by the equation $\hat{H}\ket{\Psi} = E\ket{\Psi}$, can be recast as the generalized eigenvalue problem $ HV = ESV$ within the Krylov subspace. Here, $H$ denotes the Hamiltonian matrix within the Krylov subspace, as described by Eq.~\ref{krylov_ref},
\begin{equation}\label{eq_hkl}
H_{KL} = \bra{\psi_K} 
 \hat{H} \ket{\psi_L}
= \bra{\Phi_{K}} \hat{U}^\dagger_{K} 
 \hat{H} \hat{U} _{L} \ket{\Phi_{L}}.
\end{equation}
Additionally, $S$ represents the overlap matrix among Krylov vectors, with individual elements given by,
\begin{align}\label{overlap}
S_{KL} = \langle \psi_K \ket{\psi_L}
= \bra{\Phi_{K}} \hat{U}^\dagger_{K}
\hat{U}_{L} \ket{\Phi_{L}}.
\end{align}
Considering each Krylov vector may contain distinct entanglers and reference states, the matrix elements of both $H$ and $S$ are measured on quantum computers using an ancillary qubit and the Hadamard test~\cite{Huggins_2020,jctc.9b01125}, as illustrated in Figure~\ref{fig:htest}. Once the elements of $H$ and $S$ are measured, the generalized eigenvalue problem $HV=ESV$ can be trivially solved on classical computers due to the small dimensionality of the Krylov subspace.

\begin{figure}[!htb]
  \centering
  \includegraphics[width=0.45\textwidth]{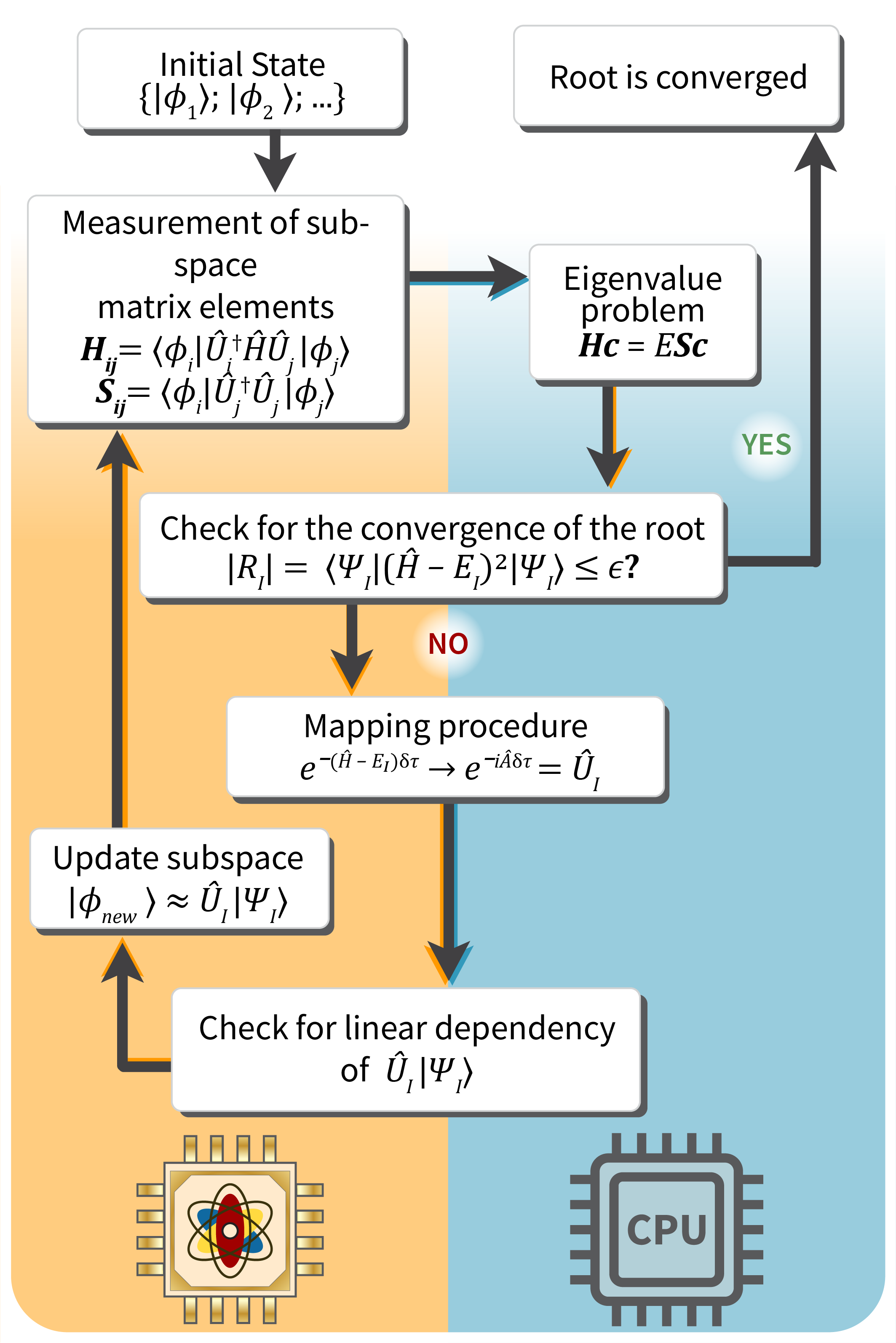}
  \caption{
  Flowchart of the QDavidson algorithm. The orange (blue) box represents the part of the algorithm that is performed on QPU (CPU).
  }
  \label{fig:flowchart}
\end{figure}

After solving for the approximate excited states in the current subspace, the residues of these approximate states can be computed as 
\begin{equation}
  \ket{R_I} = \hat{H}\ket{\Psi_I} - E_I \ket{\Psi_I} = (\hat{H} - E_I) \sum_K V_{KI} \ket{\psi_K}.
\end{equation}
The norm of the Residue can be measured on quantum computers similarly to the \(H_{KL}\) measurement,
\begin{align}
  |\ket{R_I}| = &\bra{\Psi_I} (\hat{H} - E_I)^2 \ket{\Psi_I} \nonumber\\
  = &\sum_{KL} V^*_{KI} V_{LI} \bra{\psi_K} (\hat{H} - E_I)^2 \ket{\psi_L}.
\end{align}
Hence, the measurement of $|\ket{R_I}|$ is akin to the \(H_{KL}\) measurement but with the Hamiltonian $\hat{H}$ in Equation~\ref{eq_hkl} replaced by $(\hat{H}-E^2_I)^2$. The norm indicates the convergence of each eigenstate. If $|\ket{R_I}|$ exceeds $\epsilon$ (where $\epsilon$ denotes the convergence criterion), a new Krylov vector, based on the normalized counterpart of $\ket{R_I}$, should be incorporated into the Davidson algorithm, provided it is linearly independent within the existing Krylov subspace.

The classical Davidson employs the Krylov subspace $\mK_r((\hat{H}-E),\ket{\Phi_I})$ to solve the eigenvalue problem. However, it is non-trivial to create the $\hat{H}\ket{\Psi_I}$ state on quantum circuits with the non-unitary $\hat{H}$. As an alternative, within the QDavidson algorithm, a correction vector defined by
\begin{equation}\label{eqimag1}
  \ket{\delta_I}=e^{-\Delta\tau (\hat{H}-E_I)}\ket{\Psi_I}
\end{equation}
is employed to expand the Krylov subspace. In other words, the QDavidson algorithm uses the subspace of $\mK_r(e^{-\Delta\tau(\hat{H}-E)},\ket{\Phi_I})$ to derive the eigenstates. Because $\ket{\delta_I}=[1-\Delta\tau(\hat{H}-E)]\ket{\Phi_I}+\mathcal{O}(\Delta\tau)$, it can be verified that $\mK_r(e^{-\delta\tau(\hat{H}-E)},\ket{\Phi_I})$ is equivalent to $\mK_r((\hat{H}-E),\ket{\Phi_I})$ when $\Delta\tau\rightarrow 0$.
The evolution detailed in Equation~\ref{eqimag1} is subsequently mapped to unitary operators $e^{-i\hA}$ as proposed in Ref.~\cite{GarnetNP2020},
\begin{equation}
  \ket{\delta_I} \simeq n_I e^{-i\hA}\ket{\Phi_I}
\end{equation}
where $n_I$ is the normalization factor and $\hA=\sum_\alpha a_\alpha\hP_\alpha$. The coefficients $a_\alpha$ are obtained by solving a linear algebra associated with the mapping (more details can be found in Appendix~A). An alternative method involves directly mapping the preconditioned residue $\frac{1}{H-E_i}(\hat{H}-E)$ into $e^{-i\delta\tau\hat{A}}$, represented by $\ket{\delta_I} = \frac{1}{H-E_i}(\hat{H}-E) \ket{\Phi_I} \approx Ce^{-i\hat{A}} \ket{\Phi_I}$.

After mapping the residue operator to unitaries, the QDavidson algorithm determines if the new Krylov vectors (or the correction vector) are linearly dependent within the current subspace by introducing $\ket{\delta'_{K'}}$
\begin{eqnarray}\label{delta_i2}
\ket{\delta'_{I}}&&\equiv \ket{\delta_{I}}-\sum_{KJ} \ket{\Psi_K}(S^{-1})_{KJ}\bra{\Psi_J}\delta_{I}\rangle\nonumber\\ 
&&\equiv \ket{\delta_{I}} - \sum_{KJ} (S^{-1})_{KJ} d_{IJ} \ket{\delta_I}.
\end{eqnarray}
Here, $d_{KJ}=\bra{\Psi_J}\delta_{I}\rangle$ and the norm $|\ket{\delta'_{I}}|$ is measured is measured on quantum computers. If the ratio $\frac{|\ket{\delta'_{I}}|}{|\ket{\delta_{I}}|}>\epsilon$, it implies that the new Krylov vector $\ket{\delta_I}$ is linearly independent within the current subspace, and hence, $\ket{\delta'_{I}}$ should be incorporated into the subspace. As the inclusion of a new Krylov vector into the Krylov space introduces additional correlations, the next iteration will draw the results nearer to the exact solutions. The flowchart of the QDavidson algorithm is summarized in Fig.~\ref{fig:flowchart}.

Since the size of the Krylov subspace $N_K$) is small, computing the generalized eigenvalue problems on classical computers is cheap. The primary complexity of the QDavidson algorithm stems from mapping the residue vectors into unitaries. Hence, its main bottleneck is the formation of $S$, $b$, and the resolution of the linear system to obtain unitary $\hA$. Earlier research has shown that mapping non-unitary exponential operators into unitaries scales exponentially with correlation domain $D$~\cite{GarnetNP2020}. But, a local approximation can be applied to eliminate the exponential dependence on $D$~\cite{GarnetNP2020}, leading to polynomial complexity.

\begin{figure*}[!htb]
  \includegraphics[width=0.95\textwidth]{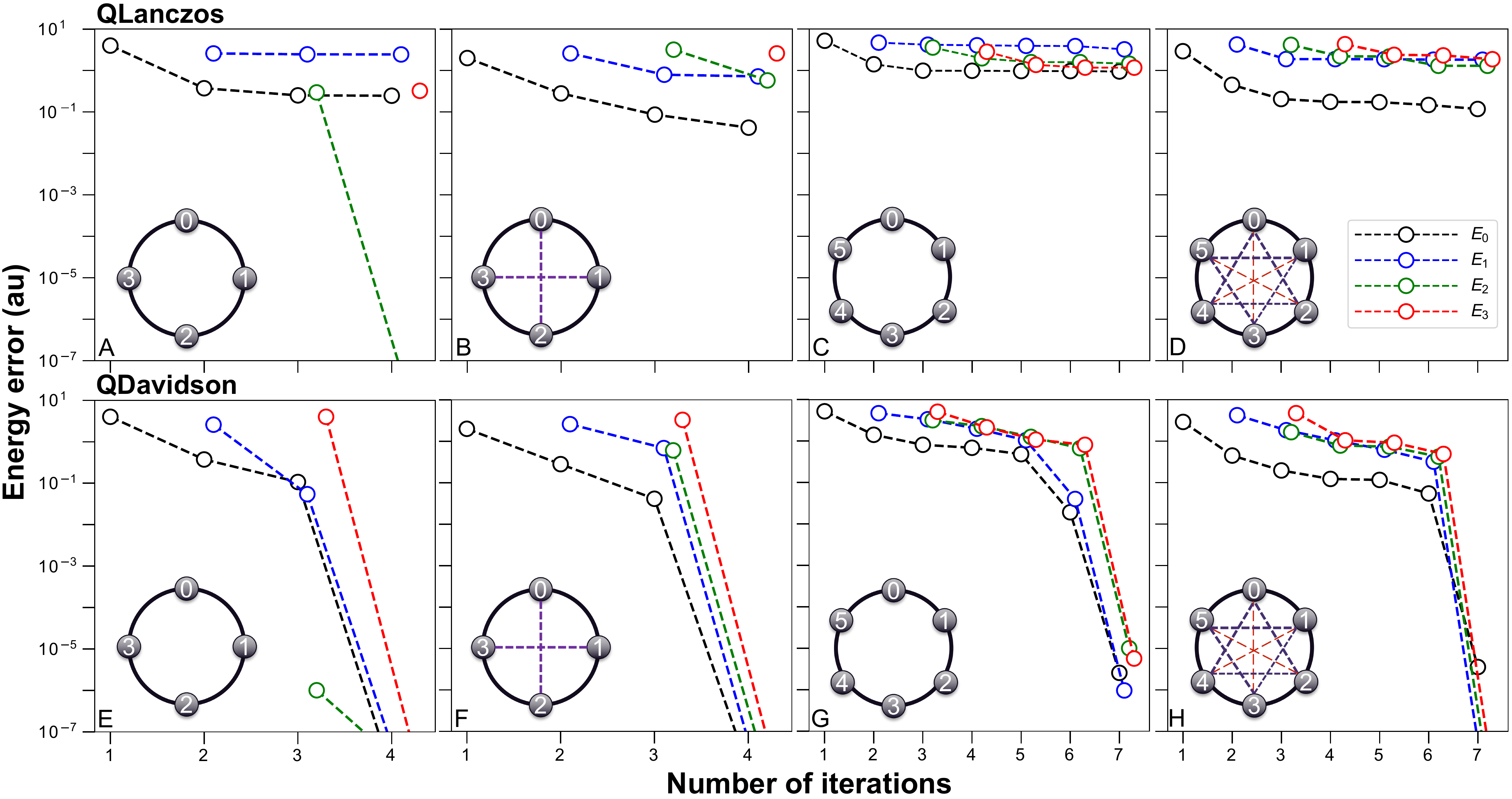}
  \caption{
  Energy differences of the 1D Heisenberg models as a function of algorithm iteration. The energy difference is defined as \(E_{\text{algorithm}} - E_{\text{exact}}\). Results for different states are represented by different colors: black for the ground state, blue for the first excited state, green for the second excited state, and red for the third excited state. Graphs (a-d) present the results of the QLanczos algorithm, while graphs (e-h) depict the results of the QDavidson algorithm obtained with the same QITE expansion parameters. Graphical representations of each system are provided in the bottom-left corner of each graph. The small circles with numbers symbolize spins, and the lines connecting these circles denote \(C_k\hat{S}_i\hat{S}_j\) terms. Different \(C_k\) coefficients are illustrated by lines of varying colors and widths.
  }
  \label{fig_davison_sr05}
\end{figure*}

\section{Numerical experiments}
In this section, we present numerical results for the proposed algorithm. To demonstrate the performance of the QDavidson algorithm, we conducted exact quantum simulations of various systems, such as one-dimensional (1D) Heisenberg models and molecular systems, using a noiseless state-vector simulator.

\subsection{One-dimensional Heisenberg models}
Both long-range (LR) and short-range (SR) one-dimensional (1D) Heisenberg models, analogous to the models described in Ref.~\cite{GarnetNP2020}, were tested in this work. The short-range models account only for nearest-neighbor interactions between spins, characterized by $C_{ij}(\hat{X}_i\hat{X}_j+\hat{Y}_i\hat{Y}_j+\hat{Z}_i\hat{Z}_j)$ terms. In contrast, the long-range models consider pairwise interactions among all spins, encompassing a larger number of terms in a qubit Hamiltonian. Explicitly, the long-range and short-range Hamiltonians are given by
\begin{equation}
  \hat{H}_{SR} = -\sum_{i=1}^{N}{\hat{S}_{i} \hat{S}_{i+1}},
\end{equation}
and
\begin{equation}
  \hat{H}_{LR} = -\sum_{i,j;j>i}^{N}{\frac{1}{D_{ij}}\hat{S}_{i} \hat{S}_{j}}, 
\end{equation}
where $D_{ij} = \min(1+|j-i|, 1+N-|j-i|)$. $N$ denotes the number of spins in the system. For the short-range Hamiltonians, the index $i$ is cyclic; thus, when it reaches $N+1$, it reverts to 1. The QDavidson algorithm is state-dependent, necessitating the definition of initial reference states. We initiated the algorithm with the anti-ferromagnetic product state, which corresponds to the alternating $\ket{01..}$ state in a computational basis. In order to benchmark the QDavidson algorithm against other QKS methods, a QLanczos algorithm with identical initial parameters was also implemented.

The results of the QDavidson and QLanczos algorithms for the 1D-Heisenberg models are shown in Figure~\ref{fig_davison_sr05}. For the initial four low-lying solutions, the QDavidson algorithm achieved convergence within 4 and 7 iterations for 4-spin and 6-spin systems, respectively (Figure~\ref{fig_davison_sr05}E-H). This performance surpasses the recently proposed QLanczos~\cite{GarnetNP2020} algorithm (Figure \ref{fig_davison_sr05}A-D), which not only converges slower but also displays numerical instability as the resultant states quickly become linearly dependent. By implementing root convergence criteria and a linear dependency check, the QDavidson algorithm exhibits enhanced numerical stability. The full results for the QLanczos algorithm are provided in the Supplementary Information (SI) (Figure~\ref{fig_qlanc_SI}). This algorithm was executed until an iteration produced a linearly dependent state. Notably, while the QLanczos managed to compute exact energy values for all 4-spin models within 6 iterations (Figure \ref{fig_qlanc_SI}A-B), it did not converge for 6-spin systems, and the final energy estimations showed a considerable deviation (Figure~\ref{fig_qlanc_SI}C-D).

\begin{figure*}[htb]
  \includegraphics[width=0.95\textwidth]{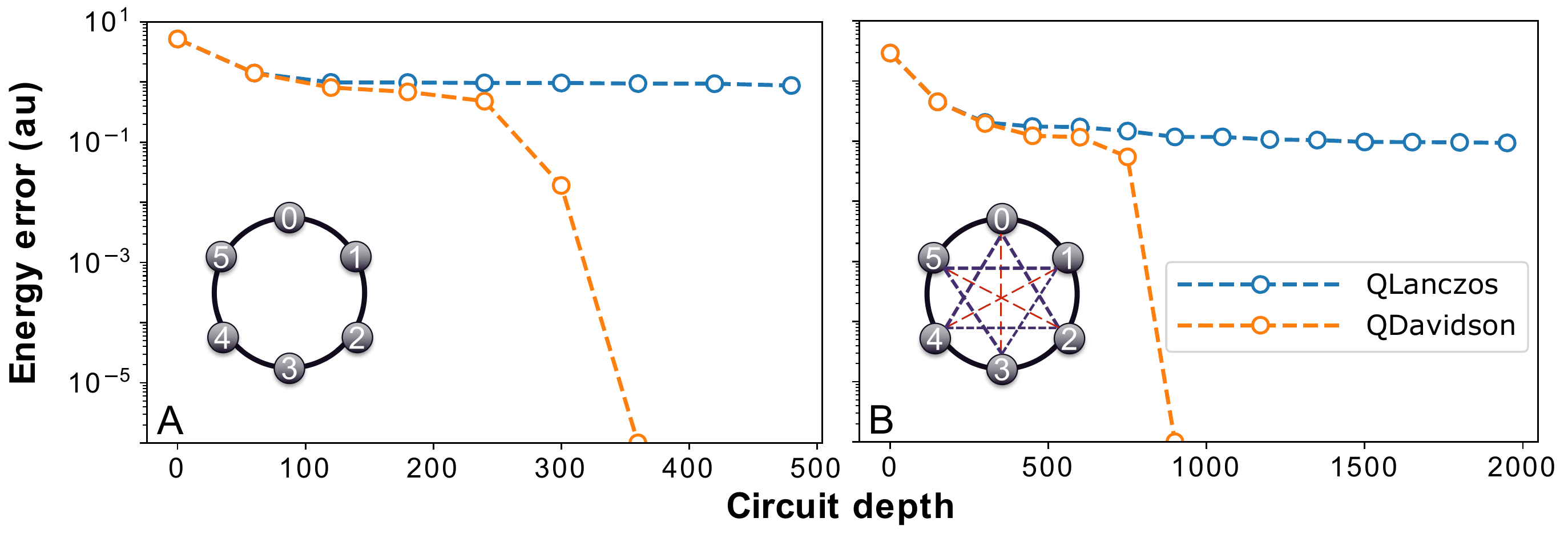}
  \caption{Energy differences of the lowest states for 6-spin 1D Heisenberg models as a function of a maximum circuit depth (A) SR model; (B) LR model.}
  \label{fig_cirq_depth}
\end{figure*}

As each iteration appends a new entangler after mapping the imaginary time evolution operator into unitaries, the circuit depth increases monotonically with iterations within the QITE, QLanczos, and QDavidson frameworks. Hence, faster convergence translates to more compact circuits. Compared to QITE algorithms, QLanczos converges faster in finding the ground state~\cite{GarnetNP2020}. Our QDavidson algorithm substantially improves convergence by employing the residue operator to narrow the subspace search near the exact state, resulting in a significantly reduced circuit depth. Although this improvement may not be evident for smaller systems, it becomes remarkably advantageous for larger systems. To elucidate this, we examined how the lowest-state energy's accuracy varies with the resulting circuits' maximum depth for both QLanczos and QDavidson. The results are shown in Figure~\ref{fig_cirq_depth}. QDavidson reproduces the exact solution for six-spin systems when the subspace comprises circuits with maximum gate depths of approximately 400 and 900 for the short-range and long-range models, respectively. In contrast, QLanczos results are less accurate, and deeper circuits are required to perform the algorithm. For instance, the accuracy attained by QLanczos after 8 iterations (gate depth of 480) for the 6-spin SR model can be achieved by QDavidson using circuits with a peak gate depth of just 120. Given the iterative expansion of the subspace, each QDavidson algorithm iteration contains states characterized by circuits of various depths. Consequently, tracking solely the longest circuit within the subspace is sufficient. The circuit depth was analyzed based on the Pauli words present in the operator $\hat{A}$ in the $e^{-\Delta\tau (\hat{H}-E)}\ket{\Psi_I} \rightarrow e^{-i\hat{A}}\ket{\Psi_I}$ expansion. Recognizing the elements of $\hat{A}=\sum_i \theta_i \hat\sigma_i$, a standard unitary evolution circuit could be constructed for $e^{-i\hat{A}}$ (assuming one Trotter step).

\begin{table}[htb]
\begin{tabular}{|l|l|l|l|}
\hline
\textbf{Molecule}  & \textbf{Initial} & \textbf{Number of} & \textbf{Energies} \\
&\textbf{state}&\textbf{iterations}& \\ \hline
$C_2H_4$ & $\ket{0011}$ & 1 & -77.11518 ($S_0$), $S_z$ = 0 \\
&&&-76.37541 ($S_2$), $S_z$ = 0 \\ \hline
&$\ket{1001}$ & 1 & -76.92218 ($T_1$), $S_z$ = 0 \\
&&&-76.58276 ($S_1$), $S_z$ = 0 \\ \hline
&$\ket{0101}$ & 0$^a$ & -76.92218 ($T_1$), $S_z$ = 1 \\ \hline
$C_3H_3^+$ & $\ket{000011}$ & 2 & -113.64929 ($S_0$), $S_z$ = 0 \\
&&& -113.19854 ($S_1$), $S_z$ = 0 \\
&&& -112.75964 ($S_2$), $S_z$ = 0 \\
&&& -112.68141 ($S_3$), $S_z$ = 0 \\ \hline
&$\ket{001001}$ & 2 & -113.35209 ($T_1$), $S_z$ = 0 \\
&&& -113.19854 ($S_1$), $S_z$ = 0 \\
&&& -112.96818 ($T_2$), $S_z$ = 0 \\
&&& -112.75964 ($S_2$), $S_z$ = 0 \\ \hline
&$\ket{000101}$& 0$^a$ & -113.35209 ($T_1$), $S_z$ = 1\\ \hline
\end{tabular}
\footnotesize{$^a$ The initial product state is already an exact solution.}
\caption{Results of QDavidson algorithm for $C_2H_4$ and $C_3H_3^+$ molecules. The initial state and number of iterations for the algorithm convergence are given. Different eigenstates are denoted in parenthesis as the following: $S_0$ - ground singlet state, $T_1$ - the lowest triplet state, $S_i$ - $i^{th}$ excited singlet state, $T_i$ - $i^{th}$ excited triplet state. The $S_z$ denotes the spin projection of the obtained state. }
\label{table_energies}
\end{table}

\begin{figure*}
  \includegraphics[width=0.95\textwidth]{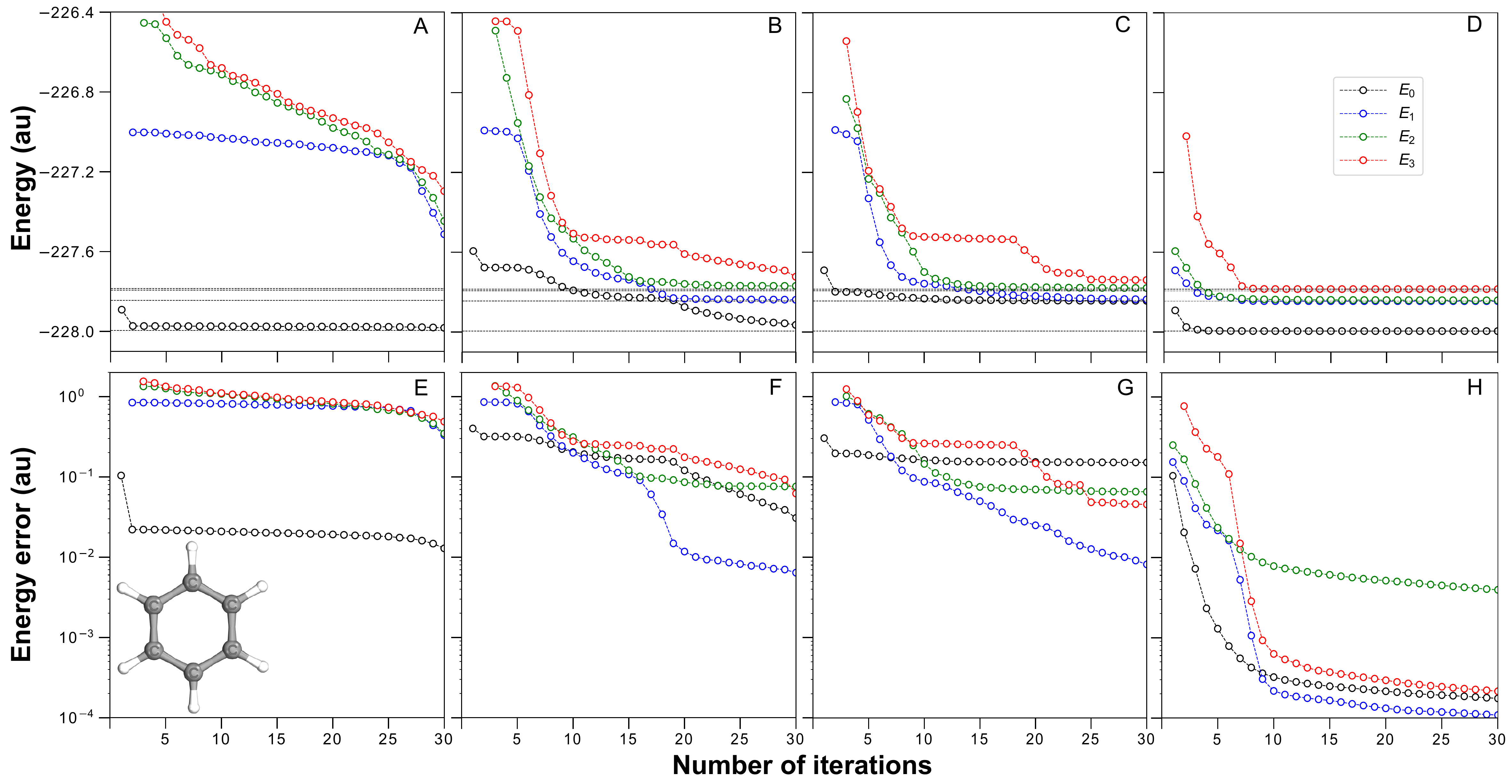}
  \caption{Energies and errors of the first four low-lying states ($S_0$, $T_1$, $S_1$, $T_2$) obtained from the QDavidson procedure as a function of algorithm iteration. Results for different states are given in different colors. The results for $C_6H_6$ molecule initialized with different product states: $\ket{000000111111}$ (A, E); $\ket{000010011111}$ (B, F); $\ket{000001011111}$ (C, G); a combination of three product states mentioned above (D, H). Black dashed horizontal lines show the exact energies obtained from a direct diagonalization of the corresponding Hamiltonian. The geometry of the molecule is shown in the inset of (E).}
  \label{fig_molecules}
\end{figure*}

\subsection{$\pi$-conjugated hydrocarbons}

In addition to the 1D Heisenberg models, we also examined our algorithm on chemical systems. As expected, molecular Hamiltonians are notably more complex than the 1D Heisenberg models due to the intricate entanglement between numerous orbitals. To highlight the advantage of the QDavidson algorithm for molecular systems, we studied three molecular systems: ethylene, cyclopropene cation, and benzene. The Cartesian coordinates of the investigated molecular systems are provided in the Supplementary Information (SI) (see Table~\ref{table_xyz}). An active space representing the $\pi$-conjugated system was selected for all the systems. The detailed active orbitals can be found in the SI (see Figure~\ref{fig_cas}). Therefore, the (2e, 2o) active space was considered for the $C_2H_4$ molecule, (2e, 3o) for $C_3H_3^+$, and (6e, 6o) for $C_6H_6$. The Jordan-Wigner transformation~\cite{JW_transformation} was utilized to map the second-quantized Hamiltonian into the qubit Hamiltonian. Consequently, the largest system examined in this study is the benzene molecule, comprising 6 electrons and 12 spin-orbitals in the active space (which corresponds to a 12-qubit Hamiltonian with 407 terms).

Taking into account that the electronic structure problems in molecular systems maintain particle-preserving and total spin-preserving symmetries, it is beneficial to establish a symmetric pool of Pauli terms for the correction vector $\ket{\delta_I}$ mapping. Therefore, a set of all possible spin projection-preserving single and double excitation operators was selected and then transformed into qubit operators employing the Jordan-Wigner scheme~\cite{JW_transformation}. All unique Pauli terms with an odd number of $\hat{Y}$ operators were selected into the pool. This results in 12, 40, and 828 unique Pauli terms for the ethylene, cyclopropene cation, and benzene, respectively.

The performance of the QDavidson algorithm depends on the defined initial state. Multiple simulations were conducted using varying initial configurations ($\ket{\phi_0}, \ket{\phi^a_i}=a^\dag_a a_i\ket{\phi_0}$). In particular, the Hartree-Fock, the first excited singlet, and the first triplet product states were taken into consideration. The QDavidson algorithm converges to exact energy values for the smallest molecular systems within several iterations. For $C_2H_4$, starting from the Hartree-Fock product state, the QDavidson algorithm obtained the exact ground state energy (-77.11518 Hartree) and the exact second excited singlet state energy (-76.37541 Hartree) after just one iteration. Similar results were acquired for the $C_3H_3^+$ system, where the exact ground state (-113.64929 Hartree) and the first three singlet excited states were determined. Table \ref{table_energies} presents results for other initial states. Intriguingly, the eigenstates identified for the $C_3H_3^+$ system maintain particle preservation and do not coincide with a neutral state solution with 3 electrons instead of two, even though they are lower in energy.

For the more extensive $C_6H_6$ system, the QDavidson algorithm also accurately located the energies of several low-lying states (within chemical accuracy). However, as depicted in Figure \ref{fig_molecules}a-c, many more iterations were needed. This is presumably due to the limited operator pool chosen, which only includes Pauli words from JW-mapped single and double excitation operators. Consequently, multi-electron systems (with electron count exceeding 2) might lack sufficient flexibility in the $e^{-\Delta\tau (\hat{H}-E)}$ operator mapping. Regardless, the algorithm yielded meaningful results even with this restricted operator pool. For the benzene molecule, the algorithm found the first excited singlet and triplet states within $10^{-2}$ Hartree after approximately 30 iterations (see Figure~\ref{fig_molecules}b, c).

For larger systems, it becomes evident that initiating the algorithm from a singular initial state isn't the optimal strategy to capture the first few low-lying excited states, as illustrated in Figure \ref{fig_molecules}A-C. Even though the algorithm can locate states of interest, the accuracy is far from the desired precision of $10^{-3}$ Hartree. Even initiating the algorithm with the first excited singlet (Figure \ref{fig_molecules}F) or triplet (Figure \ref{fig_molecules}G) determinants leads to quicker convergence compared to using the Hartree-Fock initial state. However, the energy convergence remains slow and often diverges from the exact values. In turn, the generalization of the algorithm with multiple initial states results in faster convergence and higher accuracy (see Figure~\ref{fig_molecules}H), highlighting the significance of multireference states. Notably, chemically accurate energies for $S_0$, $T_1$, and $T_2$ states could be obtained after just 11 iterations. Therefore, to efficiently find the low-lying excited states of a large molecular system, it is advised to begin with multiple initial states to expedite energy convergence.

\subsection{Effects of statistical shot-noise}
Owing to the finite number of quantum measurements (shots), exact measurements of the Hamiltonian matrix and overlap matrix elements within the Krylov subspace are unattainable. A limited shot count may introduce numerical instabilities into the QDavidson procedure. In general, to estimate the expectation value of a Hamiltonian with respect to a single state $\ket{\psi}$ with precision $p$, it is required to perform $O(|h_{max}|^2 Mp^{-2})$ measurements, where $h_{max}$ is the largest coefficient in the Hamiltonian decomposition $\hat{H}=\sum_i{h_i\hP_i}$ \add{, } and $M$ denotes the number of Hamiltonian terms~\cite{RN155}. Similar conclusions apply when evaluating Hamiltonian matrix elements within the Krylov subspace. Since $\bra{\Psi_I} \hat{H} \ket{\Psi_J} = \sum_i{h_i\bra{\Psi_I} \hP_i \ket{\Psi_J}} = \sum_i\sum_{KL}V^*_{KI}V_{LJ}\bra{\psi_K} \hP_i\ket{\psi_L}$ and each element $\bra{\psi_K} \hP_i\ket{\psi_L}$ can be evaluated according to Figure~\ref{fig:htest}, the cost of evaluating the Hamiltonian matrix element is $O(|h_{max}|^2 Mp^{-2}N^2_K)$ where $N_K$ is a size of the Krylov subspace. 
Although this measurement process can be complex, in practice, only a small subspace is typically spanned, and the growth of the subspace is linear with respect to the number of iterations, making the $N^2$ factor relatively insignificant compared to $|h_{max}|^2 Mp^{-2}$. Likewise, the overlap matrix's measurement incurs a cost of $O(p^{-2}N^2_K)$ shots. The procedure for mapping residual vectors to unitaries also requires quantum measurements. The required shot count here depends on the size of the operator pool ($P$) chosen for the mapping and is in order of $max[O(|h_{max}|^2 Mp^{-2}N^2_K P),O(p^{-2}N^2_K P^2)]$.
However, it is noteworthy that the shots needed for QDavidson to converge are fewer than for the VQE procedure since the latter's ansatz optimization involves a substantial number of energy evaluations.

To illustrate the number of shots required to perform the algorithm on a real-world example, we incorporated the shot-noise into the calculations for the $C_2H_4$ molecule. The desired precision $p$ was determined after examining the algorithm's noise robustness by introducing the random error to each element of Hamiltonian matrix, (off-diagonal) overlap matrix element, $S_{\alpha\beta}$ and $b_{\alpha}$ (Appendix A). It was found that an evaluation precision of $10^{-4}$ is sufficient to guarantee the algorithm's numerical stability. With this level of precision, energy values of $-77.115\pm0.001 (S_0)$ and $-76.375\pm0.001 (S_2)$ were replicated. The algorithm, when initiated with a Hartree-Fock reference state, converged after 1 iteration. The estimated shot count per circuit evaluation was set to $10^8$, and the total number of shots was found to be $\sim10^{10}$. This experiment underscores the QDavidson algorithm's capability to reproduce the low-lying spectra of the evaluated Hamiltonian even in the presence of statistical shot noise.

\section{Summary}
In this study, we developed an efficient QKS algorithm, QDavidson, to compute ground states and low-lying excited states by harnessing the Krylov subspace's power and the Davidson algorithm's rapid convergence. Unlike other QKS methodologies that employ real or imaginary time to pre-generate a subspace, QDavidson uses residues from previously approximated states to expand the subspace and capitalizes on the pre-conditioner to narrow the subspace search near the exact state, ensuring rapid convergence. Our numerical simulations confirm that the QDavidson algorithm surpasses other QKS methods like QLanczos in terms of convergence speed. Since the circuit depth increases with iterations, QDavidson's rapid convergence results in less complex quantum circuits, enhancing its resilience against noise. Future research could delve into advanced pre-conditioners for the QDavidson method~\cite{jctc6b00459,davidson2021shane}, potentially further trimming the iterative steps and circuit depth. Moreover, simulation accuracy is bound by the chosen basis set. While a large basis set is essential for achieving chemical accuracy, increasing the basis set size in quantum algorithms is not NISQ friendly, which requires a significantly larger number of qubits and deeper circuits~\cite{transcorrH2022}. However, by employing a trans-correlated Hamiltonian, one can attain accuracy at the cc-pVTZ basis even with a minimal basis set~\cite{transcorrH2022}.

\begin{acknowledgements}
The research presented in this article was supported by the Laboratory Directed Research and Development (LDRD) program of Los Alamos National Laboratory (LANL) under project number 20200056DR. We thank the LANL Institutional Computing (IC) program for access to HPC resources. ST acknowledges the support from the Center of Integrated Nanotechnologies (CINT), a US Department of Energy and Office of Basic Energy Sciences User Facility. LANL is operated by Triad National Security, LLC, for the National Nuclear Security Administration of the US Department of Energy (contract no. 89233218CNA000001). AIB and NVT acknowledge funding from the subcontract with LANL (Award ID: 203369-00001). AIB acknowledges financial support from the R. Gaurth Hansen Professorship Fund.
\end{acknowledgements}

\bibliography{ref}

\appendix
\clearpage 

\begin{center}
  {\bf Supplementary Information for ``Quantum Davidson Algorithm for Excited States"}
\end{center}
\renewcommand\thefigure{S\arabic{figure}}
\renewcommand\thetable{S\arabic{table}}
\renewcommand\theequation{S\arabic{equation}}
\renewcommand\thealgorithm{S\arabic{algorithm}}

\setcounter{figure}{0}
\setcounter{table}{0}
\setcounter{algorithm}{0}
\setcounter{equation}{0}

\section{Mapping Residue into unitaries}

In this appendix, we discuss the way of mapping Residue in the Davidson algorithm, $(\hat{H}-E_i)\ket{\Phi_i}$, into unitaries $e^{-i\hA}{\Phi_i}$ on quantum circuits.
Because 
\begin{equation}
  \hat{H}-E_i=\frac{1-\delta\tau(\hat{H}-E_i+1/\delta\tau)}{\delta\tau}\simeq\frac{e^{-\delta\tau(\hat{H}-E_i+1/\delta\tau)}}{\delta\tau},
\end{equation}
The Davidson generator can be obtained by using an energy shift of $E-1/\delta\tau$ in the exponential operator. Define $\hat{H}'_i=\hat{H}-E_i+1/\delta\tau$
\begin{equation}
  \ket{\Phi'}=c^{1/2}e^{-\delta\tau\hat{H}'_i}\ket{\Phi_i}\simeq e^{-i\hA\delta\tau}\ket{\Phi_i}. 
\end{equation}
where $\hA=\sum_\alpha a_\alpha\hP_\alpha$ and $\hP_\alpha$ are the Pauli words. $c$ is the normalization factor. The above mapping can be achieved by using the method developed in Ref.~\cite{GarnetNP2020}. The coefficients $a_I$ can be solved via the linear algebra $Sa=b$ where
\begin{equation}
  S_{\alpha\beta}=\bra{\Phi_I}\hP^\dag_\alpha\hP_\beta\ket{\Phi_I}
  =\sum_{KL}V^*_{KI}V_{LI}\bra{\psi_K}\hP^\dag_\alpha\hP_\beta\ket{\psi_L}.
\end{equation}
and 
\begin{equation}
  b_\alpha=-\frac{i}{\sqrt{c}}\Im[\bra{\Phi_I}\hP^\dag_\alpha\hat{H}'\ket{\Phi_I}].
\end{equation}

\begin{algorithm}[H]
\caption{Classical Davidson algorithm for obtaining $N_I$ lowest eigenvalues (and eigenvectors) of a matrix $H$ with initial orthogonal guess vector $\{\ket{b_I}\}$ ($\bra{b_I}b_J\rangle=\delta_{IJ}$).}\label{alg_dav0}
\begin{algorithmic}[1]
\Procedure{DAVIDSON}{$H, N_I, \{\ket{b_I}\}$} 
\State Form the subspace matrix $\bar{H}=\bra{b_I}H\ket{b_J}$.
\State Solve the subspace eigenvalue problem: $\bar{H}_{IJ}C_{JK}=C_{IK}h_K$.

\State Obtain the $N_I$ lowest eigenvalues $h_K$ and the corresponding eigenvectors $\ket{\Phi_I}=\sum_K C_{KI}\ket{b_K}$.
\State Form the residual vectors: $\ket{r_I}=(H-E_I)\ket{\Phi_I}$.
\State Check the norm $|\ket{r_I}|$ of the residues for the convergence of each root norm($r_I)\leq \epsilon$?. 
\For{I in the non-converged roots}
\State Form the pre-conditioned correction vector:
$\ket{\delta_I}=-(\bar{H}-h_I)^{-1}\ket{r_I}$ and normalize.
\State Orthogonalize to the current trial subspace:
$\ket{\delta'_I}=\ket{\delta_I}-\sum_J\ket{b_J}\bra{b_J}\delta_I\rangle/\bra{b_J}b_J\rangle$.
\If{norm($\ket{\delta'_I}$)/norm($\ket{\delta_I}$)$>\epsilon$}
  \State Normalize $\ket{\delta'_I}$ and add it to the subspace.
\EndIf
\EndFor
\EndProcedure
\end{algorithmic}
\end{algorithm}

\begin{figure*}[htb]
  \includegraphics[width=0.95\textwidth]{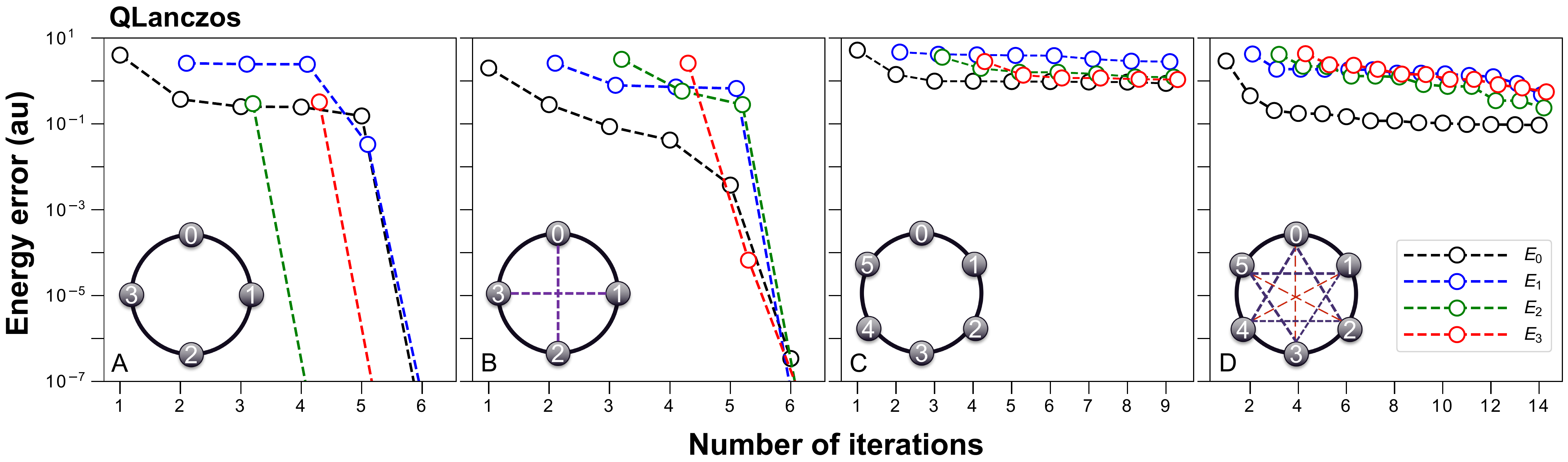}
  \caption{Energy differences of the 1D Heisenberg models~\cite{GarnetNP2020} as a function of QLanczos algorithm iteration. The energy difference is defined as $E_{algorithm}-E_{exact}$. Results for different states are given with different colors: black - ground state; blue - first excited state; green; second excited state; red - third excited state. Graphical representations of each system are given in the bottom left corner of each graph. The small circles with numbers represent spins, while lines connecting those circles represent $C_k\hat{S}_i\hat{S}_j$ terms. Different $C_k$ coefficients are illustrated with lines of different colors and widths.}
  \label{fig_qlanc_SI}
\end{figure*}

\begin{figure*}[htb]
  \includegraphics[width=0.95\textwidth]{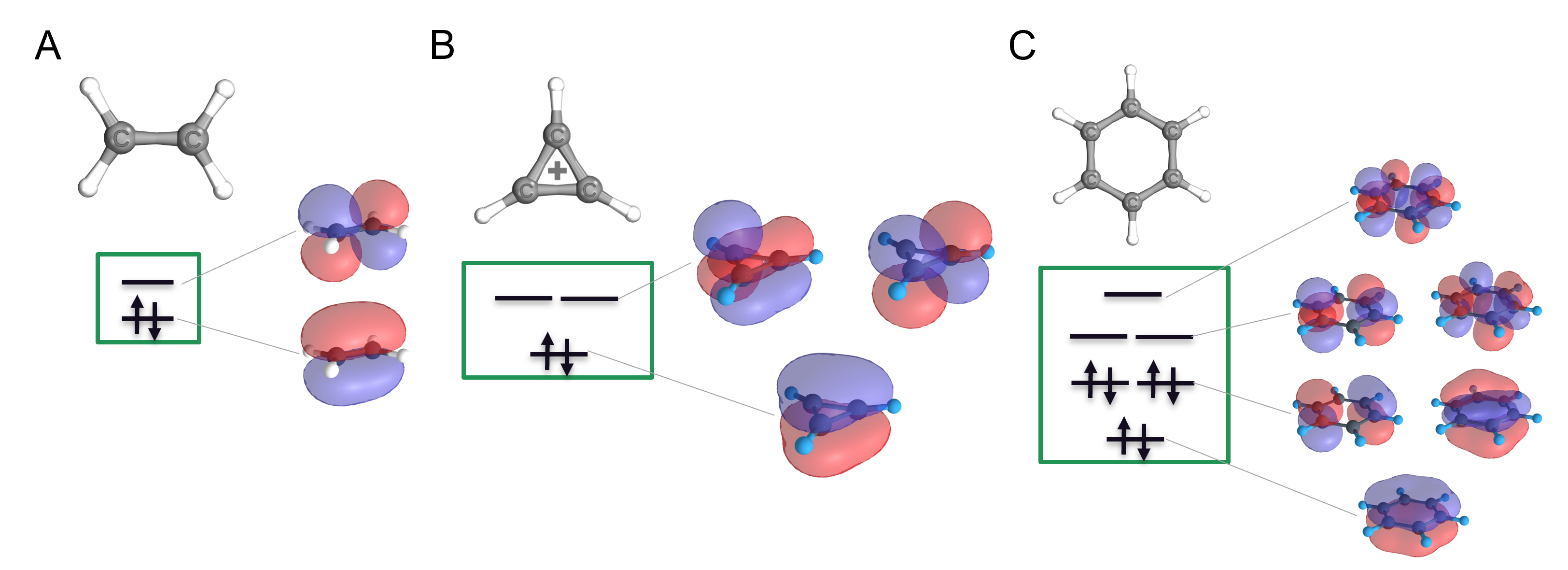}
  \caption{
  Molecular orbitals chosen into an active space of investigated molecules: (A) ethylene; (B) cyclopropene cation; (C) benzene; (D) cyclooctatetraene dication.
  }
  \label{fig_cas}
\end{figure*}

\begin{table}[H]
\begin{tabular}{|l|l|}
\hline
$C_2H_4$                    & Basis set: STO-3G \\& Hartree-Fock Energy: -77.0739546295 Hartree \\ \hline
                             & C        0.000000000      0.000000000      0.652999000
\\&C        0.000000000      0.000000000     -0.652999000
\\&H        0.000000000      0.915997000      1.229260000
\\&H        0.000000000     -0.915997000      1.229260000
\\&H        0.000000000      0.915997000     -1.229260000
\\&H        0.000000000     -0.915997000     -1.229260000
                                                                                                                \\ \hline
$C_3H_3^+$ & Basis set: STO-3G \\& Hartree-Fock Energy: -113.620321176  Hartree          \\ \hline
                             &  C        0.000000000      0.794870000      0.000000000
\\&C        0.688378000     -0.397435000      0.000000000
\\&C       -0.688378000     -0.397435000      0.000000000
\\&H        1.636710000     -0.944955000      0.000000000
\\&H        0.000000000      1.889910000      0.000000000
\\&H       -1.636710000     -0.944955000      0.000000000
                                                                                                               \\ \hline
  $C_6H_6$                   &       Basis set: STO-3G\\&  Hartree-Fock Energy: -227.891360223 Hartree   \\ \hline
                             &      C        0.000000000      1.386832000      0.000000000
\\&C        1.201032000      0.693416000      0.000000000
\\&C       -1.201032000      0.693416000      0.000000000
\\&C        1.201032000     -0.693416000      0.000000000
\\&C       -1.201032000     -0.693416000      0.
\\&C        0.000000000     -1.386832000      0.000000000
\\&H        2.138466000      1.234644000      0.000000000
\\&H       -2.138466000      1.234644000      0.000000000
\\&H        2.138466000     -1.234644000      0.000000000
\\&H       -2.138466000     -1.234644000      0.000000000
\\&H        0.000000000     -2.469288000      0.000000000
\\&H        0.000000000      2.469288000      0.000000000
                                                                                                           \\ \hline

\end{tabular}
\caption{Cartesian coordinates of molecular systems that were used in this study.}
\label{table_xyz}
\end{table}

\end{document}